\documentclass[manuscript]{aastex}

\usepackage{fancyhdr}
\usepackage{longtable}
\usepackage{latexsym}
\usepackage{graphicx}
\usepackage{amsmath}
\usepackage{natbib}
\usepackage{upgreek}
\usepackage{multirow}
\usepackage{array}
\usepackage{pdflscape}
\usepackage{graphicx,subfigure}
\usepackage{tabularx}
\usepackage{epsfig}
\usepackage{color}
\usepackage{epstopdf}
\usepackage{geometry}

\newcommand{\tabincell}[2]{\begin{tabular}{@{}#1@{}}#2\end{tabular}}
\newcommand{\PreserveBackslash}[1]{\let\temp=\\#1\let\\=\temp}
\newcolumntype{C}[1]{>{\PreserveBackslash\centering}p{#1}}
\newcolumntype{R}[1]{>{\PreserveBackslash\raggedleft}p{#1}}
\newcolumntype{L}[1]{>{\PreserveBackslash\raggedright}p{#1}}

\shorttitle{Flaring M Dwarfs in the LAMOST-Kepler Field}
\shortauthors{H.-Y. Chang et al.}

\begin{document}

\title{LAMOST Observations of Flaring M Dwarfs in the Kepler Field}

\author{H.-Y. Chang \altaffilmark{1},
       Y.-H. Song  \altaffilmark{2},
       A.-L. Luo \altaffilmark{2},
       L.-C. Huang \altaffilmark{1},
       W.-H. Ip \altaffilmark{1,3},
       J.-N. Fu \altaffilmark{4},
       Y. Zhang \altaffilmark{5},
       Y.-H. Hou \altaffilmark{5},
       Z.-H Cao  \altaffilmark{2},
       Y.-F  Wang \altaffilmark{5}
}

\altaffiltext{1}{Institute of Astronomy, National Centrel University, , 300 Jhongda Road, Jhongli 32001, Taiwan. 
\sf wingip@astro.ncu.edu.tw}
\altaffiltext{2}{Key Laboratory of Optical Astronomy, National Astronomical Observatories, Chinese Academy of Sciences, Beijing 100012, China. }
\altaffiltext{3}{Space Science Institute, Macau University of Science and Technology, Avenida Wailong , Taipa, Macau. }
\altaffiltext{4}{Department of Astronomy, Beijing Normal University, 19 Avenue Xinjiekouwai, Beijing 100875, China.}
\altaffiltext{5}{Nanjing Institute of Astronomical Optics \& Technology, National Astronomical Observatories, Chinese Academy of Sciences, Nanjing 210042, China}

\begin{abstract}
A sample of the LAMOST spectra of the early type M0-M3 dwarfs is compared with the Kepler observations. It is found that M dwarfs with strong chromospheric emission in $H_{\alpha}$  have large flare activity in general. The rotational periods derived from the Kepler measurements have close correlations with the sizes of the flares, the power-law distribution index and the equivalent widths of the $H_{\alpha}$ emission. A clear trend exists for higher magnetic activities being detected in faster rotating M dwarfs (rotation periods $<$ 20 day).

\end{abstract}


\section{Introduction}

Solar and stellar flares are explosive events because of the sudden release of a large amount of energy in a short time interval. The occurrences of solar flares and the related coronal mass ejection events are topics of great interest and have been subject to intensive efforts of monitoring observations in different wavelengths. In the case of the stellar flares, the study of their physical characteristics and time evolution is, however, hampered by limitation in time coverage and sample size intrinsic to ground-based observations. These constraints have been partially elevated by the dedicated space-based monitoring observations provided by the Kepler satellite (Borucki et al., 2010). In addition to the search of exoplanets from the un-interrupted measurements of the light curves of 150,000 stars in white light between 4300 \AA $ $ -- 8900 \AA, a wealth of data on variable stars and stellar flares has been archived for public use. The round-the-clock monitoring of a large number of stars over a time interval of nearly 4 years overcomes the difficulties encountered by ground-based observations suffered from frequent interruptions in the long-term time-series photometric measurements. This point has been dramatically exemplified by the discovery of superflares of solar-like stars by Maehara et al. (2012) and the analysis of the flare activity of a number of A-F stars by Balona (2012). 

Because of the strong magnetic activity associated with their convective envelopes, the M dwarfs (dM stars) are known to have frequent flares, especially in the case of the late-type stars (i.e., M3 -- M9) according to Hilton (2011). UV Ceti-type flares have been well-documented even though whether the generation mechanism is similar to that of solar flares is still uncertain (Moffett, 1974; Dal and Evren, 2012). Because the evolution of the M dwarfs are closely relevant to the dM stars interesting issue of exoplanetary habitability (Tarter et al., 2007), it is therefore interesting to investigate their flare phenomena.  The Kepler data are ideal for this purpose. 

Hawley et al. (2014) and Davenport et al. (2014, 2015) have examined the flare activity (or inactivity) of a few dM stars including GJ 1243, GJ 1245AB, GJ 4083 (inactive), GJ 4099 (inactive), at short (1-min) cadence in the Kepler data. These works showed that, as in the case of the solar flares, the dM stars flares can be generally classified into two types, namely, the simple ones characterized by a single peak, and the complex ones with multiple peaks. The time evolution of the single-peak events can be further divided into three phases: the impulsive rise, impulsive decay, and gradual decay. This behavior is typical of solar flares which are generated by the conversion of magnetic energy stored in the active regions to plasma kinetic energy. This also means that the flare activity of dM stars could be correlated with the level of chromospheric emission (Silvestri et al., 2005). It is therefore important to examine whether the dM stars with high $H_{\alpha}$ emission brightness can also produce strong flare activity. The observational program of the Large Sky Area Multi-Object Fibre Spectroscopic Telescope (LAMOST) has targeted a large number of dM stars with 1800 mid-resolution spectra released in DR1 (Guo et al., 2015). Of these, 54 with good quality LAMOST spectra are in the Kepler field. These LAMOST spectra thus provide an excellent opportunity to study the nature of the dM flares by combining them with the Kepler light curves.

In this paper, we first introduce in Section 2 the LAMOST spectra of high quality with counterparts in the Kepler observations. Section 3 provides information on the flare activity of dM stars in the Kepler field with high-quality LAMOST spectra. A comparison of the Kepler and LAMOST results are given in Section 4. A summary and discussion are included in Section 5. 

\section{The LAMOST spectra}
\label{sect:lamost}
The LAMOST telescope is a quasi-meridian reflecting Schmidt telescope with an effective aperture of 4 m and a field of view of 5 degrees in diameter. It was designed to carry out spectral survey for northern sky stars, galaxies and QSOs, which can reach to -10 degrees declination (Cui et al. 2012). The field of view of LAMOST varies with sky pointing and is in average 5 degree in which 4000 fibers are applied to collect target light simultaneously feeding to 16 spectrographs. The spectral resolution is R=1800 over the entire wavelength range of 3690 -- 9100\AA\ . Two 4Kx4K CCD camera are used in the spectrograph, one for the blue range of 3690 -- 5900\AA\  and the other for the red range of 5700 -- 9100\AA\ . A one-year pilot survey was performed from October 2011 to June 2012, to be followed by a five-year regular survey started in September 2012. The scientific focus of the regular survey is mainly on Milky Way stars covering the magnitude range from 9 --17.8 mag in the r band (Luo et al. 2015). 

In the first LAMOST data release (DR1; Luo et al. 2015), the spectra of 93619 different dM stars have been obtained. Out of 110321 M dwarf spectra, 7179 indicated that the dM stars are magnetically active according to the measurements of the $H_{\alpha}$ emission (Guo et al., 2015). A subset has been further found to display variable $H_{\alpha}$ emission as indicated by observations made at different epochs. The corresponding data set in subsequent DR3 provides even more dM spectra. We have matched the LAMOST dM stars targets with those with Kepler light curves, 54 can be identified. Table 1 is a sample of their KID numbers and physical parameters derived from Huber et al. (2014) and the LAMOST pipeline. (See Appendix for the full Table 1.) 

Figure 1 shows examples of the LAMOST spectra of dM stars with varying degrees of $H_{\alpha}$ emission. KID 5791720 displays strong $H_{\alpha}$ line emission at 6562.8 \AA \ indicating significant level of stellar activity. It is followed by KID 8507979 and KID 11495571. KID 6286466 has no significant $H_{\alpha}$ emission. Note that significant levels of $H_{\beta}$ (4861 \AA) and $H_{\gamma}$(4341\AA) emission can be seen in the spectra of KID 5791720 and KID 850979 but not the other two. In fact, few spectra in this sample exhibit $H_{\beta}$ and $H_{\gamma}$ emission because of low signal noise ratio, these two emission line are located in the blue arm of LAMOST spectrograph, which has have lower throughput than red arm especially for red stars such as M type.

In order to examine the magnetic activity of different dM stars from their $H_{\alpha}$ emission, we compute the total flux of the line emission in terms of the equivalent width (EW) corresponding to the case of the line absorption profile. In this terminology, the emission EW $(H_{\alpha}$) has positive value and the absorption EW $(H_{\alpha}$) has negative value. Following the computational method of Yi et al. (2014), the EW value can be defined as, 

\begin {equation}
EW=\int (\frac{F_\lambda}{F_{O}}-1)\,d\lambda
\end {equation}

where $F_{\lambda}$ is the spectral flux at wavelength and $F_{O}$ is the average flux of the continuum background. We use a wavelength region of total width of 14 \AA  \  for the calculation of EW of $H_{\alpha}$. The central wavelength is 6564.66 \AA\ in vacuum with 7 \AA\ on either side. The continuum regions are 6555.0 -- 6560.0 \AA $ $ and 6570.0 -- 6575.0 \AA, respectively. From this procedure it can be estimated that EW = 4.42 for KID 5791720 with strong $H_{\alpha}$ emission, to be followed by EW = 3.78, 2.60, and -0.57, for KID 8507979, KID, 11495571, and KID 6286466, respectively.  

Figure 2 shows the histogram of the distribution of the EW values of the 54 dM stars (see Table 1). It can be seen that the EW distribution is not symmetrical around EW = 0. Instead, it is characterized by an excess of EW $(H_{\alpha})>$ 0. Some of them have values EW $(H_{\alpha})>$ 1 with KID 5791720 displaying the largest EW of 4.5. We have further examined the time-series spectra of those dM stars with EW $(H_{\alpha})>$ 1 and found the indication of time variability in some of them which in turn suggests the possibility of flare activity.

Another way to evaluate the level of magnetic activity of stars is to calculate the S-index dealing with the strength of the chromospheric Ca II H and K emission lines at 3934 \AA \ and 3969 \AA \ (Zhao 2011). Figure 3 shows the scatter plot of EW $(H_\alpha)$ vs. the S-indices. In the figure,  we picked out those spectra with signal noise ratio (SNR) $>$ 6 in g band from these 54 dM stars. \\

In the next Section, we will examine the flare effect in the Kepler light curves.

\section{Kepler light curves}
\label{sect:Kepler}
For the study of the random process of sudden energy release over a short time scale ($\sim$ hours) from stellar atmospheres, long-term time-series observations at comparatively short cadence ($\leq$30 min) are necessary.  From this point of view, a very valuable data archive is provided by the Kepler mission with uninterrupted monitoring of the brightness variabilities of as many as 150000 stars from 2009 to 2014 (Borucki et al. 2010). Most of the light curve measurements are for the solar-like stars according to the Kepler catalog.  If we limit the effective temperature ($T_{eff}$) range from Huber et al. (2014) to be between 2500 K and 3800 K, 3130 M dwarfs can be identified. A comparison with the LAMOST DR3 category shows that 54 Kepler dM stars have LAMOST spectra of good quality (see Table 1).

Figure 4 is a collection of short segments of the light curves of KID 5791720, KID 8507979, KID 11495571 and KID 6286466. Except for KID 6286466, all of them show flare activities. The periodic variations are presumably produced by the presence of star spots and hence represent the rotation periodicities. It also means that flares can occur at different phases of the rotational modulation of the light curves produced by star spots. To quantify the brightness variation caused by the flare events, the following procedure is applied (Wu et al., 2015). That is, we first calculate the median flux $<F>$ of a light curve over a number of periods and then generate the residuals at successive data points by computing $\Delta F=F-<F>$. The magnitude of the brightness variation is normalized to $<F>$. Figure 4.a.-- 4.c. show that the maximum amplitudes $(\Delta F/<F>)$ of the flares of both KID 5791720 and KID 8507979 can reach as much as 25$\%$ to be followed by 15$\%$ of KID 11495571. For KID 5791720, an analysis of its light curve over the four years' interval indicates that the $\Delta F$ value at the peak can actually be as high as 93$\%$ of the average stellar luminosity. Figure 5 shows one such event. The duration of the energy release is about 3.5 hours. On the other hand, there is no sign of flares in the light curve of KID 6286466 in Fig.4.d.  

The time profile of a flare allows us to estimate its total energy content which is computed by integrating the flux increment $\Delta F$ in the time interval between the starting point and the ending point when the flux level reaches the background (see Figure 5). The flare energy so obtained can be further re-normalized by multiplying it with the stellar luminosity computed by using the radius $(R_{*})$ and effective temperature ($T_{eff}$) of the stellar type. That is, flare energy $E = 4 \pi R_{*}^{2}\sigma_{sb}  T_{eff}^{4} \int \Delta F dt$ where $\sigma_{sb}$ is the Stefan-Boltzmann constant. For example, in the case of the flare event of KID 5791720 illustrated in Figure 4, the total energy in optical emission detected by Kepler can be estimated to be 7.6x10$^{35}$ ergs. Note that energy release of this range is quite comparable to the high end of the superflares of solar-type stars studied by Maehara et al. (2012). On the other hand, the energy of the superflares in Maehara et al. (2012) is at most a few percents of the luminosity of the solar-type stars themselves. 

The light curves of many of the Kepler dM stars have enough numbers of flares so that their flare frequency distributions (FFDs) can be constructed. Instead of binning the data into several energy intervals, the Maximum Likelihood method (Newman,2005) has been applied to analysis of the FFDs to extrapolate power law distributions in this study, the power law index (k) of the corresponding cumulative FFD turns out to be a useful parameter for intercomparison. 

Figure 6 shows examples of the cumulative FFDs for dM stars with frequent flare activity. 
It is noted that the FFDs tend to taper off at lower energy. This could be due to a real physical effect, namely, the occurrence frequency of small flares could become less frequently. Alternatively, the flattening of the FFDs could be caused by the limitation in counting small flares of short duration ( $<$ 30 min). Davenport et al. (2015) analyzed the short-cadence ( 1 min) light curves of GJ 1243 (M4) - namely, KID 9726699 - and produced a FFD with a power-law index of k $\sim$ 1.01 for energy between 10$^{30}$ and 10$^{33}$ ergs.  This value is consistent with the value of k $\sim$ 1.25 $\pm$ 0.19 derived from our treatment of the long-cadence data for energy between $10^{33}$ and $10^{34}$ ergs (see Figure 6.d). This intercomparison means that the flattening of the FFD of GJ 1243 pointed out before is mainly the result of low time resolution in the long-cadence light curves.

Figure 7 shows the histogram of the power-law index. The k values vary between 0.75 and 2.0 (see Table 1). While the average value is k = 1.30 $\pm$ 0.14, the variance is significant. This effect is similar to the variation of the power-law index of the FFDs of solar-type stars (Wu et at., 2015). In the following section we will examine the correlations among individual parameters associated with the flare phenomena of dM stars.

\section{Comparison}
\label{sect:comparison}
The magnetic activity of a star could be related to the strength of the dynamo mechanism which in turn could be coupled to the stellar rotation itself (Reiners and Basri 2007, 2008; Engle and Guinan, 2011). It is therefore interesting to examine the relationship between the rotation periods (P) as determined from the periodicities of the light curves and the corresponding flare activities. Within the context of the present study, we show the results of the dM stars listed in Table 1. 

Figure 8.a shows that there is an apparent cutoff at P$\sim$ 20 days for flares with large amplitude ($> 3\%$) to show up. In the case of the flare frequency distributions, the trend is for the dM stars with rotation periods $<$ 20 days to have smaller k-values.  

Figure 8.b shows the statistical correlation between the rotation period and the power law index. Interestingly, fast-rotating (or younger) dM stars appear to have harder FFDs, namely, smaller k values. This point is certainly worth further investigation by using a larger sample of dM stars.

Figure 9 is a comparison of the rotation periods with EW of $H_{\alpha}$ emission. There is a clear  feature for enhanced chromospheric activity in fast rotating dM stars. Those with P $>$ 20 days usually are inactive, at least, as indicated by the LAMOST measurements of the EW values. 

We can further examine the spectroscopic signature of flaring dM stars by producing scatter plots of the EW values against $(\Delta F/ <F>)$ and against the k-values of the corresponding FFDs. Figure 10.a shows comparison between the EW of the $H_{\alpha}$ emission and the flare amplitude $(\Delta F/ <F>)$. There is an obvious trend with a strong correlation between the maximum magnitudes of the flares of individual dM stars with their EWs. As expected, $\Delta F/<F>$ is smaller than 10$\%$  for EW $<$1. 

In Figure 10.b we can find a probable trend for the power law index value to be proportional to EW, namely, dM stars with larger EW (and hence larger flare amplitude) appear to have larger k-values (i.e. steeper FFDs) -- if only data points with EW$>$1 are considered. The reason is simply that for dM stars with small EW values and hence less flares, the statistics of the FFDs will be less accurate. Large fluctuation in the k value can therefore occur.

\section{Summary and Discussions}
\label{sect:summery}
From a match of the list of the M dwarfs (M0 -- M3) observed by Kepler and the LAMOST DR3 catalog of the M dwarf spectra, we have identified 54 M dwarfs for intercomparison. An analysis of the equivalent width of the $H_\alpha$ emission shows that a significant fraction of the dM stars are magnetically active. Those with large EW values tend to exhibit strong flare activity with some of them possessing explosive energy comparable to the luminosity of the central stars. The size frequency distributions of the flare energies of individual dM stars can be fit to power-law distributions with the power-law distribution index estimated to be 1.30 $\pm$ 0.14 on the average.   Our study also shows that fast rotating dM stars with periods $<$ 20 days are characterized by higher occurrence frequency of large flares. 

The present statistical study demonstrates the value and importance of intercomparison of the time-series observations of Kepler and the spectroscopic measurements of LAMOST. In future, we will carry out LAMOST observations of more Kepler dM stars to allow more comprehensive understanding of the magnetic behaviors of this intriguing population of low-mass stars.

\section*{Acknowledgements}

We would like to thank the Reviewer for useful comments and suggestions. The work at NCU is partially supported by grant number NSC 102-2112-M-008-013-MY3 and NSC 101-2111-M-008-016 from the Ministry of Science and Technology of Taiwan and gran number 017/2014/A1 and 039/2013/A2 of FDCT, Macau.The work at NAOC is partially supported by the National Key Basic Research Program of China (Grant No.2014CB845700), and the National Science Foundation of China under Grant Nos. 11390371, 11233004)

Guoshoujing Telescope (the Large Sky Area Multi-Object Fiber Spectroscopic Telescope, LAMOST) is a National Major Scientific Project built by the Chinese Academy of Sciences. Funding for the project has been provided by the National Development and Reform Commission. LAMOST is operated and managed by the National Astronomical Observatories, Chinese Academy of Sciences.

\clearpage

\begin{figure}
\centering
\includegraphics[width=12cm]{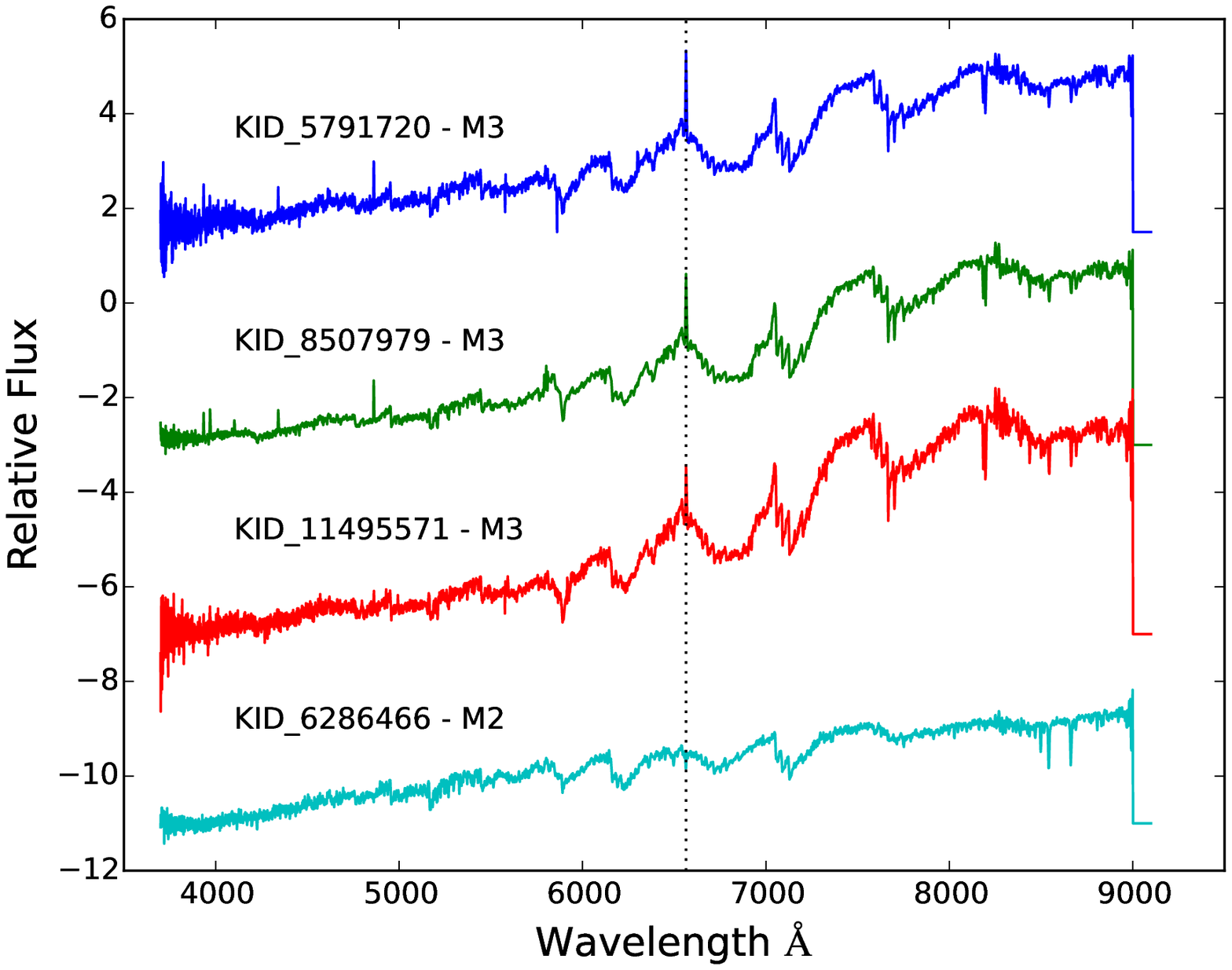}
\caption{The LAMOST spectra of four M dwarfs observed by Kepler: (a) KID 5791720; (b) KID 8507979; (c) KID 11495571; and (d) KID 6286466. In the case of KID 5791720 and KID 8507979, the $H_{\beta}$ emission at 4861 \AA \ and $H_{\gamma}$ at 4341 \AA \ can be clearly identified.}
\label{fig1}
\end{figure}

\clearpage
\begin{figure}
\centering
\includegraphics[width=12cm]{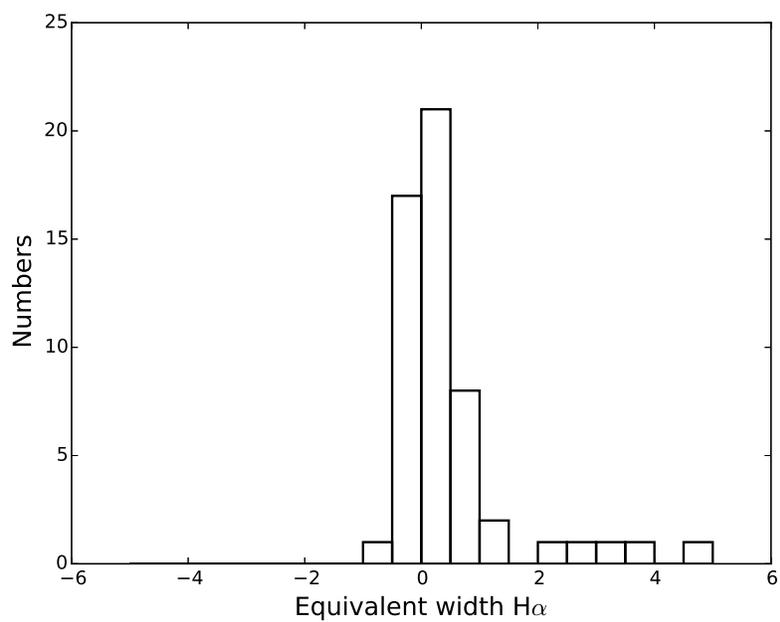}
\caption{A histogram of the EW of the $H_{\alpha}$ emissions of 54 M dwarfs with common data coverage of LAMOST and Kepler. The emission EW has positive value and the absorption EW has negative value.}
\label{fig2}
\end{figure}

\clearpage
\begin{figure}
\centering
\includegraphics[width=12cm]{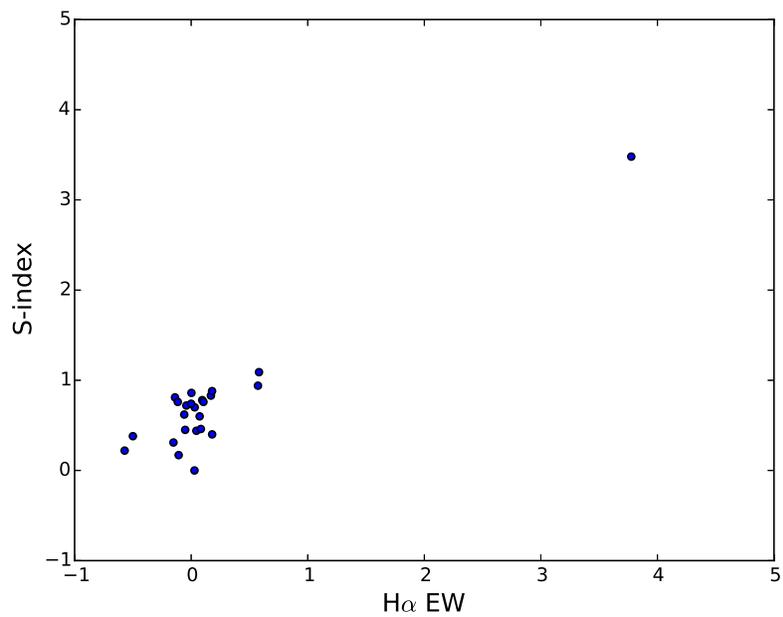}
\caption{A comparison of the EW of the $H_{\alpha}$ emission with the S-index values. Note that the SNR of S-index are larger than 6.}
\label{fig3}
\end{figure}

\clearpage
\begin{figure}
\includegraphics[width=16.0cm]{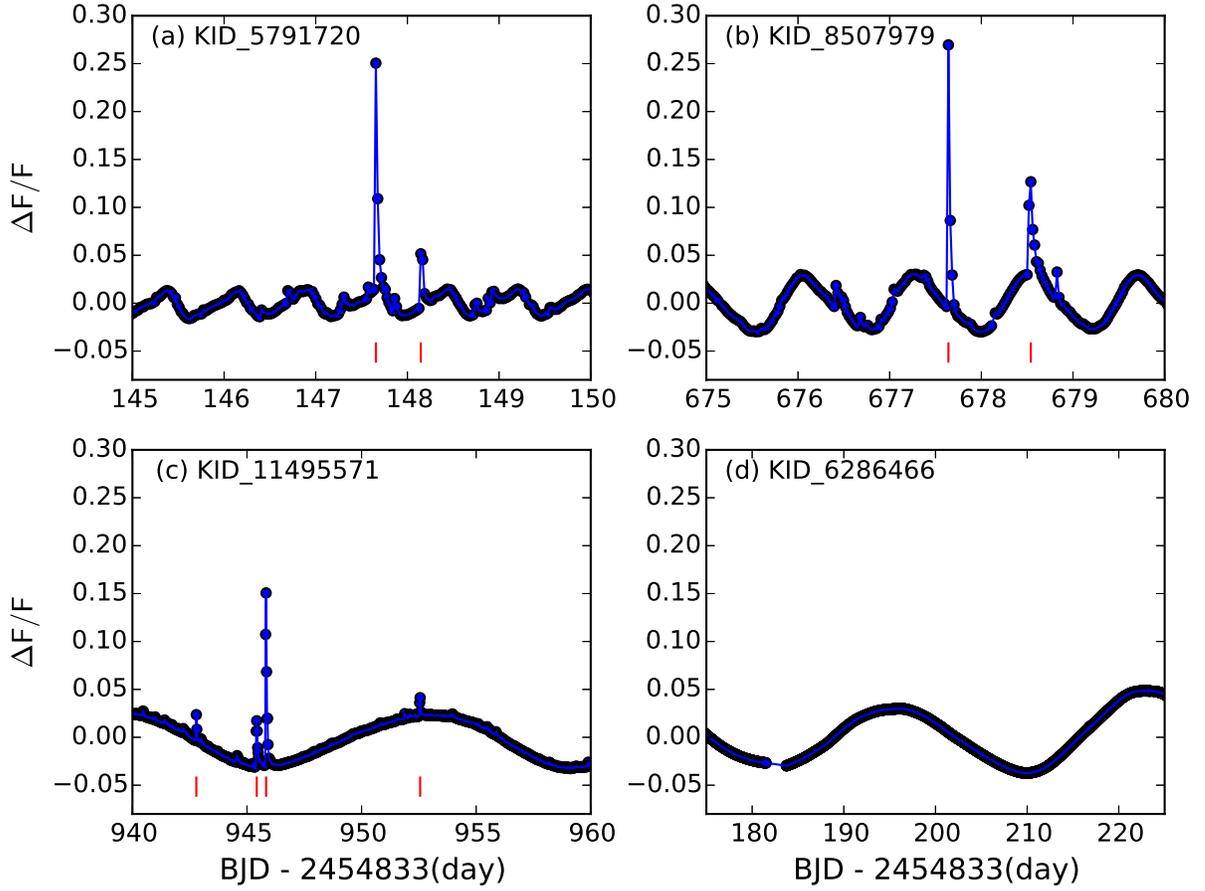}
\caption{The light curves of the dM stars: (a) KID 5791720; (b) KID 8507979; (c) KID 11495571; and (d) KID 6286466. The flares can be found at different phases of the periodic modulations of the light curves produced by star spots.}
\label{fig4}
\end{figure}

\clearpage
\begin{figure}
\centering
\includegraphics[width=12.0cm]{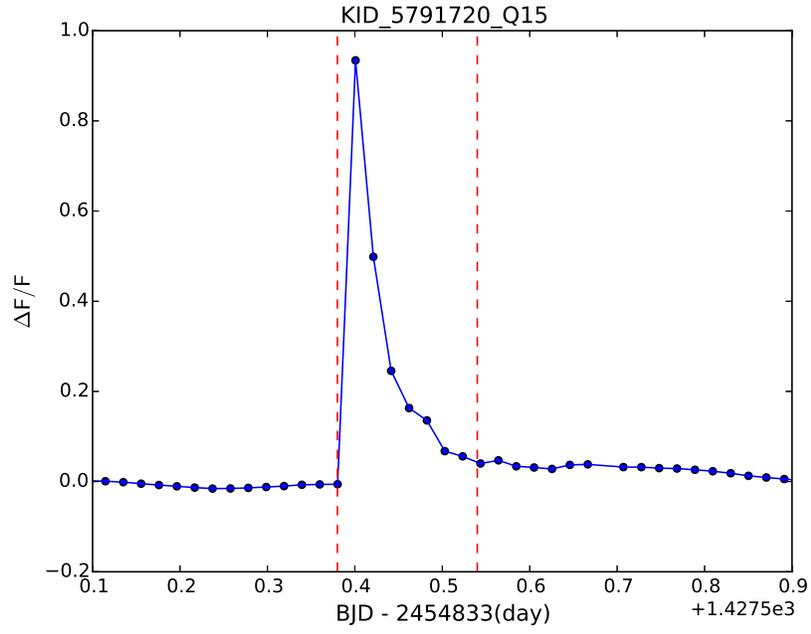}
\caption{A time profile of a super- (or hyper-) flare of KID 5791720 with the peak flux reaching almost the same level as the stellar luminosity of this M3 star. The two red dashed lines indicate the time interval for the flare energy calculation.}
\label{fig5}
\end{figure}

\clearpage
\begin{figure}
\centering
\includegraphics[width=6.5cm]{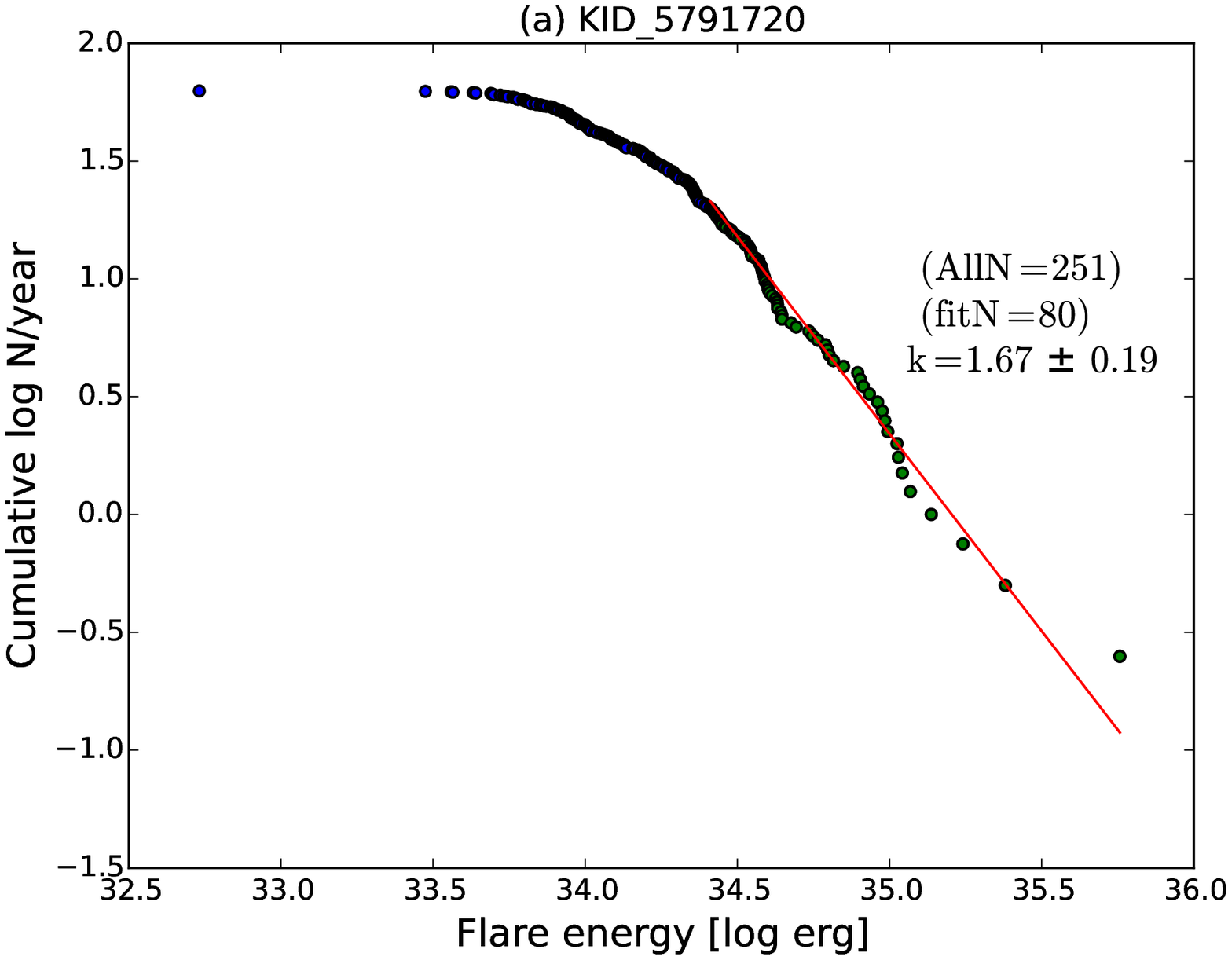}
\includegraphics[width=6.5cm]{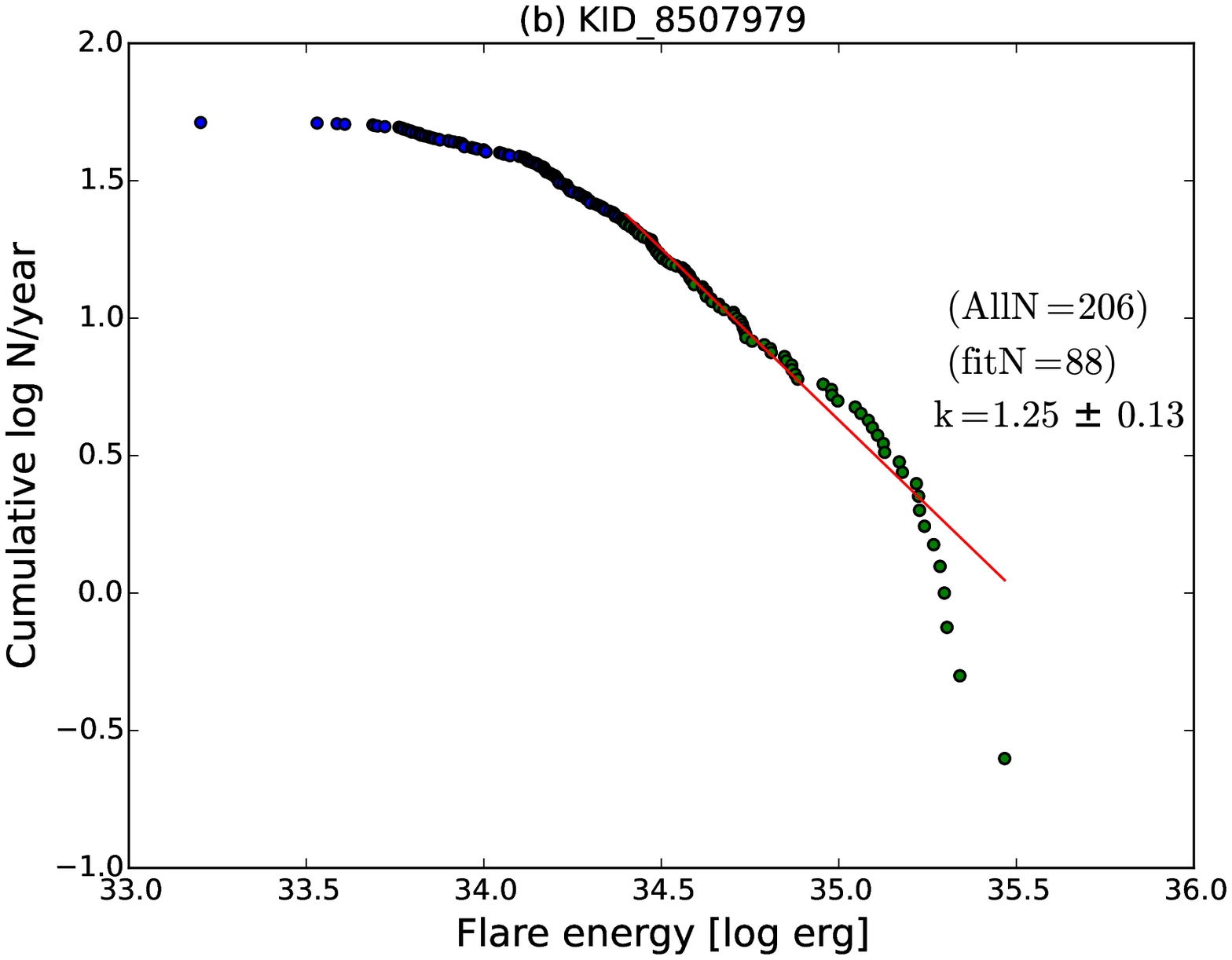}
\includegraphics[width=6.5cm]{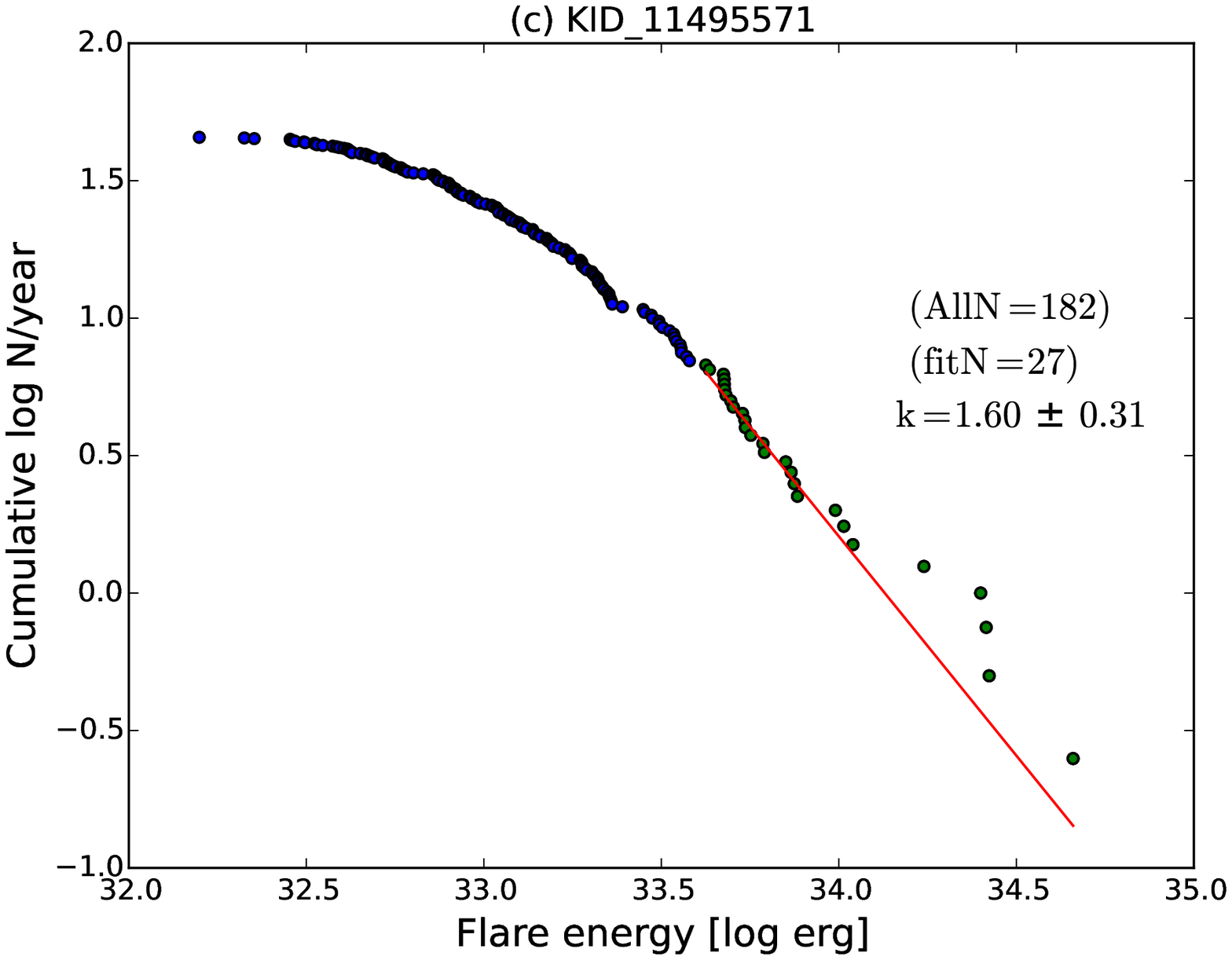}
\includegraphics[width=6.5cm]{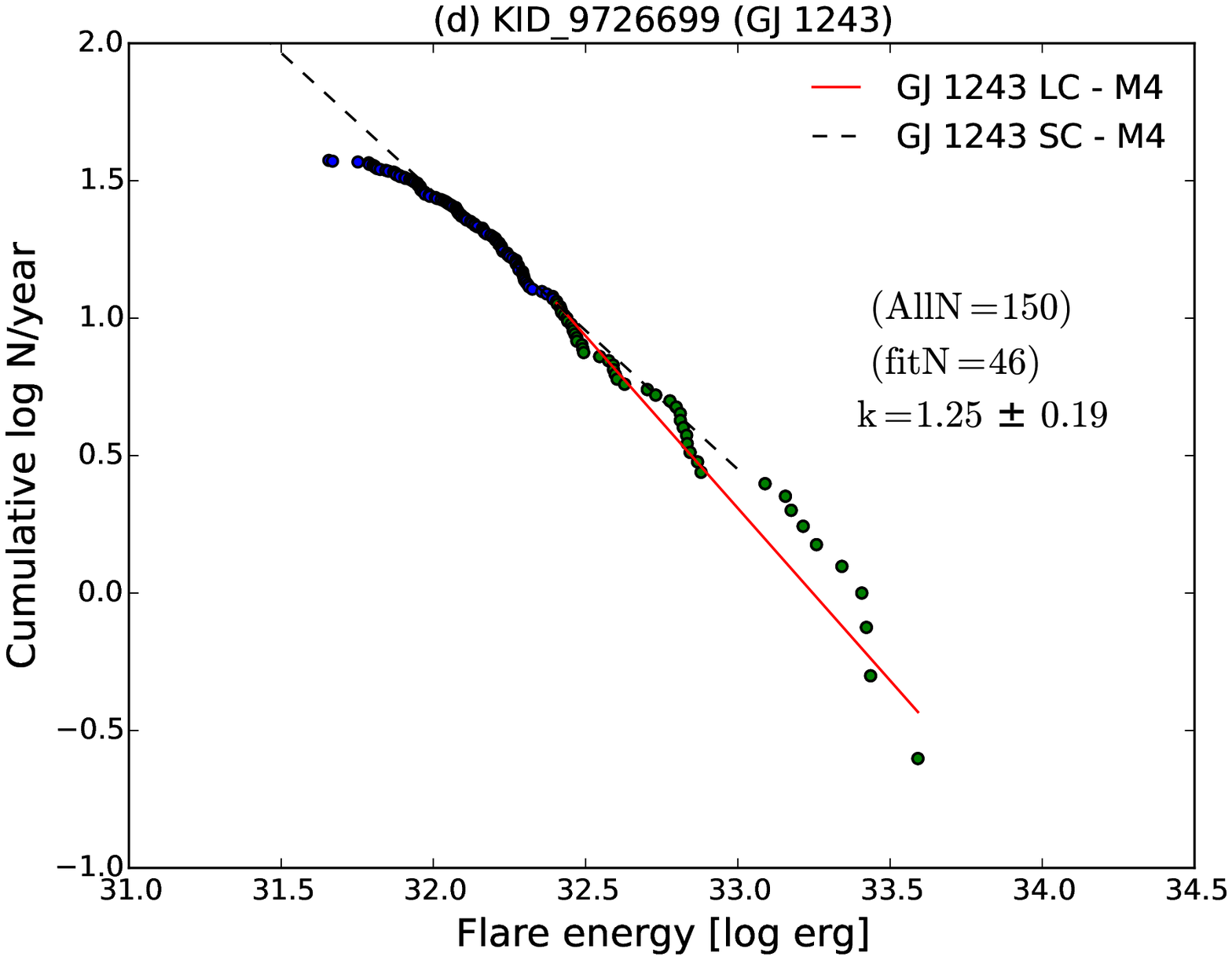}
\caption{The flare frequency distributions of four dM stars from Kepler observations: (a) KID 5791720; (b) KID 8507979; (c)KID 11495571; and (d) KID 9726699.  The power law distribution index are computed by using the Maximum Likelihood method.}
\label{fig6}
\end{figure}
\clearpage

\clearpage
\begin{figure}
\centering
\includegraphics[width=15.0cm]{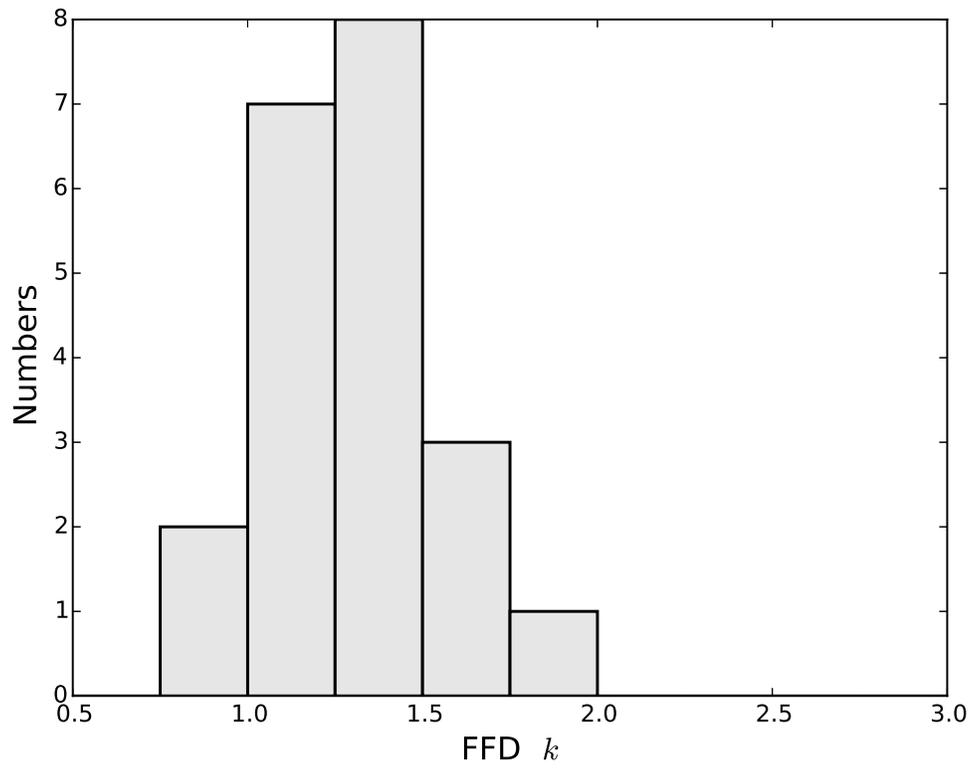}
\caption{ A histogram of the power law index (k) value.}
\label{fig7}
\end{figure}
\clearpage

\clearpage
\begin{figure}
\centering
\includegraphics[width=8.0cm]{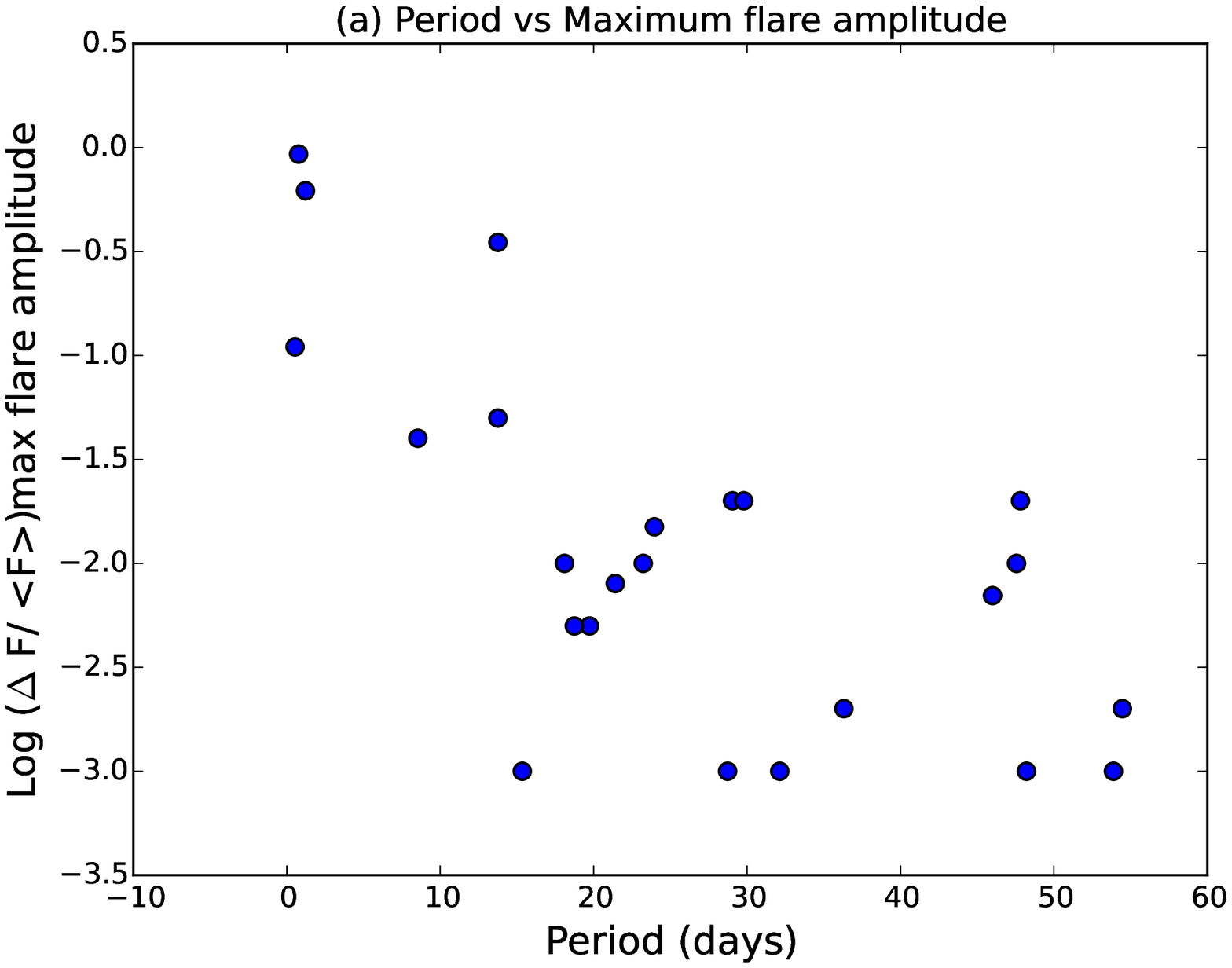}
\includegraphics[width=8.0cm]{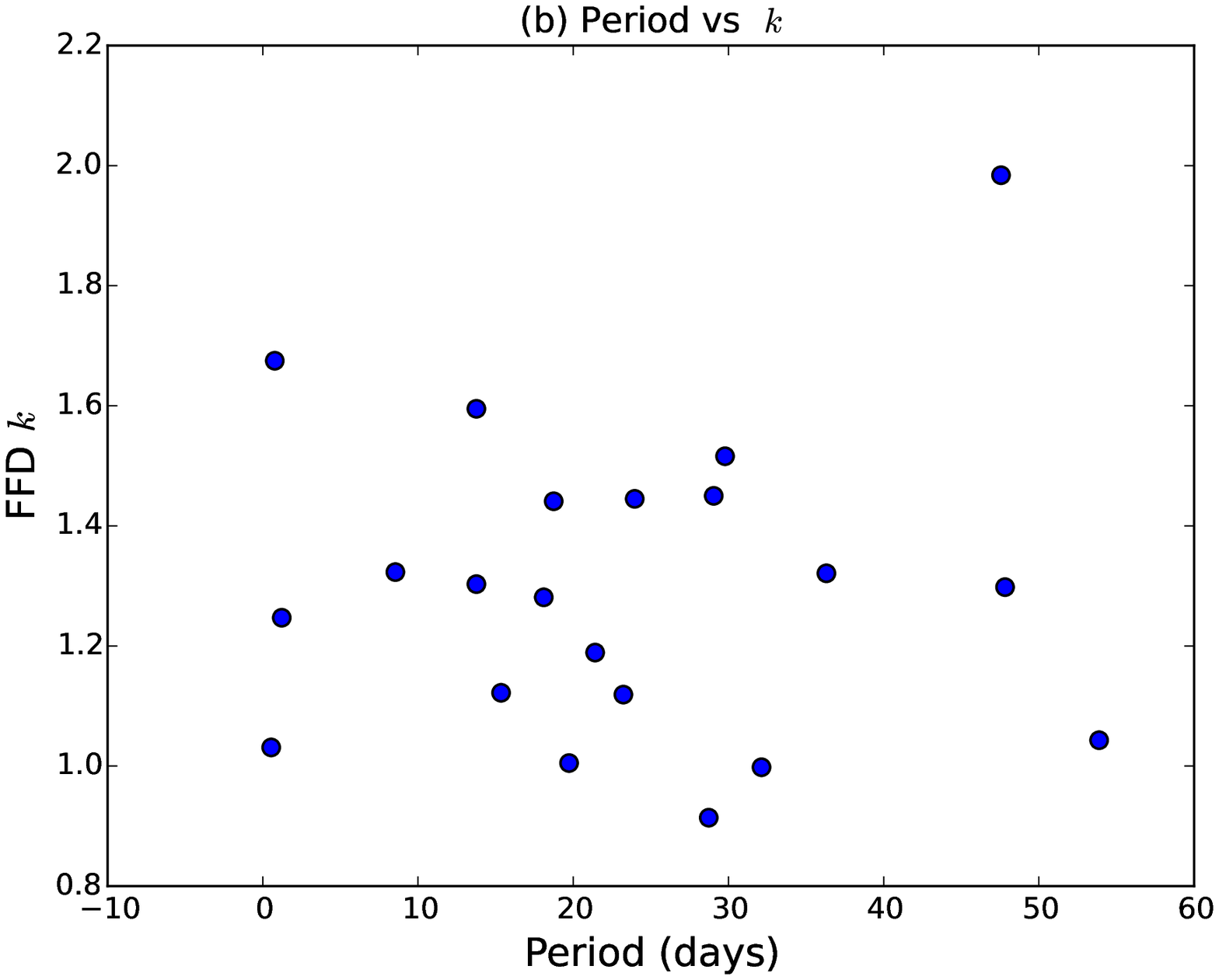}
\caption{A comparison of the rotation period with (a) the maximum flare magnitude; and with (b) the power-law distribution index. }
\label{fig8}
\end{figure}
\clearpage

\clearpage
\begin{figure}
\centering
\includegraphics[width=15.0cm]{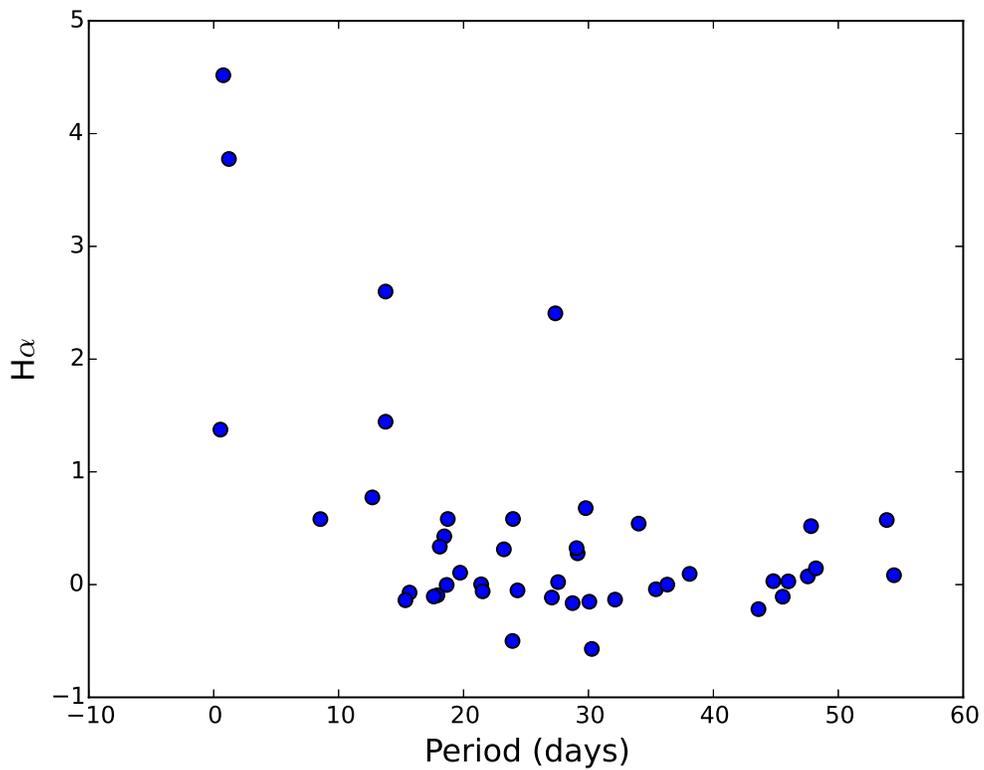}
\caption{An intercomparison of the EW values of $H_{\alpha}$ measured by LAMOST with the rotation periods of dM stars from Kepler.}
\label{fig9}
\end{figure}
\clearpage

\begin{figure}
\centering
\includegraphics[width=8.0cm]{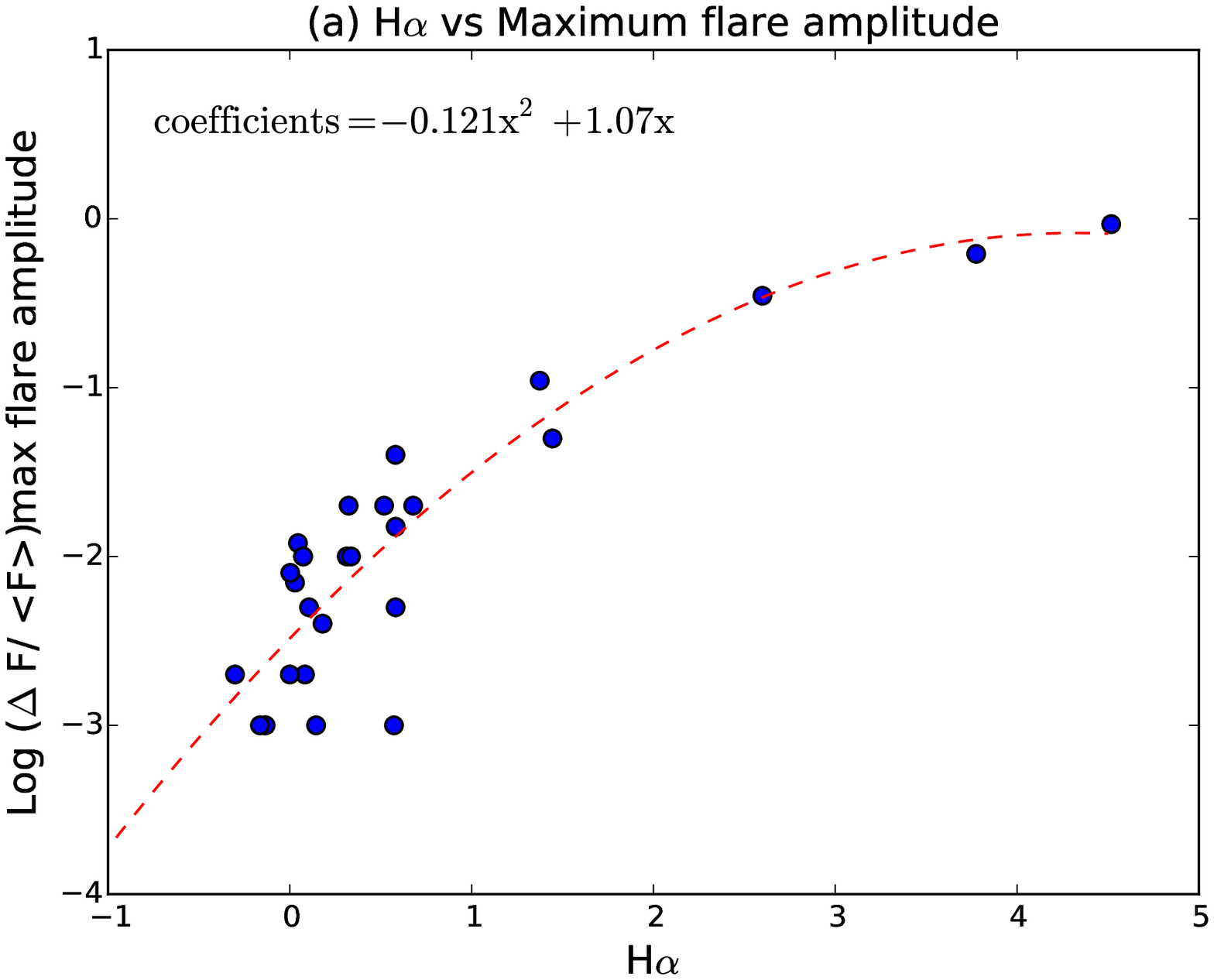}
\includegraphics[width=8.0cm]{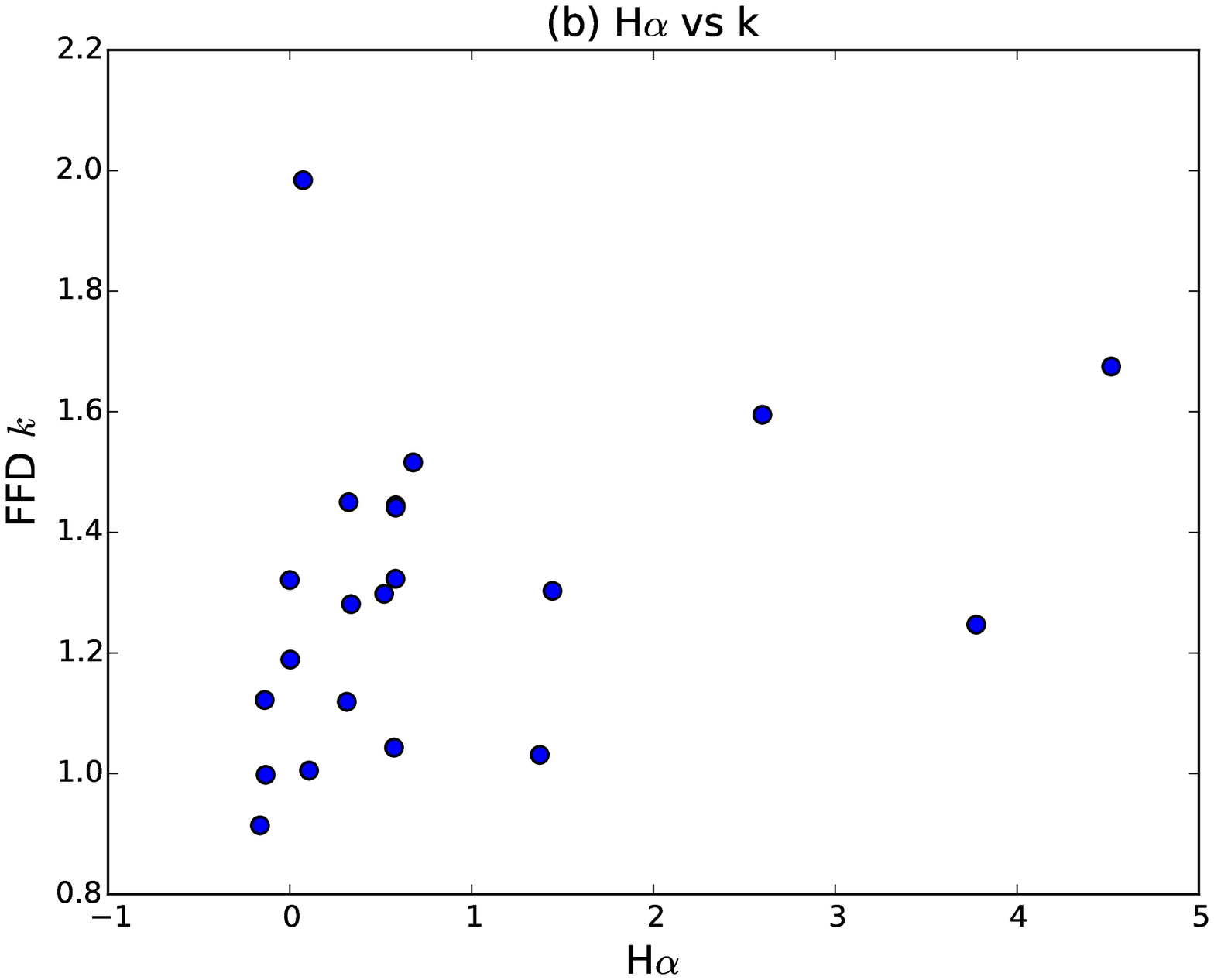}
\caption{A intercomparison of the EW values of $H_{\alpha}$ emission measured by LAMOST with (a) the maximum flare amplitude, and (b) the power-law index of the FFDs of dM stars from Kepler. }
\label{fig10}
\end{figure}

\clearpage
\newgeometry{left=1cm}

 \begin{longtable}{lllllllllll}

 \caption{List of the Kepler M dwarfs with LAMOST spectra listed in DR3.}  \label{grid_mlmmh} \\

 \hline
 \hline

  Kepler ID & $T_{eff}$$^{a}$ & \tabincell{c}{$R_{stra}$$^{a}$ \\(in R$\odot$)} & \tabincell{c}{$P_{rot}$$^{b}$ \\(day)} & Type$^{c}$  & \tabincell{c}{EW$^{d}$ \\(H$\alpha$)} & S-Index$^{e}$ & \tabincell{c}{$\Delta$F/$<$F$>$$^{f}$\\(Max)} & FFD k$^{g}$ & $N_{all}$$^{h}$  & $N_{fit}$$^{i}$  \\
 \hline

 \hline
 \hline

    1721911 & 3705  & 0.45  & ---   & M0    & 0.17  & 0.83  & ---   & ---   &       &  \\
    1866535 & 3754  & 0.473 & ---   & M0    & 0.18  & 0.88  & ---   & ---   &       &  \\
    2574201 & 3688  & 0.44  & 29.76 & M1    & -0.15 & 0.31 & ---   & ---   &       &  \\
    3438817 & 3651  & 0.496 & 19.01 & M0    & 0.09  & 0.78 & ---   & ---   &       &  \\
    3439791 & 3754  & 0.473 & ---   & M0    & 0.04  & 0.44 & 0.06  & ---   &       &  \\
    3550668 & 3477  & 0.39  & 23.44 & M1    & 0.03  & 0 &0.007 & ---   &       &  \\
    3748918 & 3747  & 0.47  & 25.38 & M0    & 0.02  & --- & ---   & ---   &       &  \\
    3946894 & 3658  & 0.45  & 19.74 & M0    & 0.11  & 0.76 & 0.005 & 1.01$\pm$0.07  & 234   & 219 \\
    4489051 & 3574  & 0.4   & 32.79 & M0    & 0.54  & ---  & --- & ---   &       &  \\
    4551429 & 3546  & 0.39  & 18.13 & M1    & -0.04 & 0.72 & ---   & ---   &       &  \\
    4770707 & 3469  & 0.315 & 41.7  & M1    & -0.22 & ---  & --- & ---   &       &  \\
    5266549 & 3704  & 0.45  & 15.72 & M0    & -0.07 & ---  & ---   & ---   &       &  \\
    5297821 & 3679  & 0.37  & ---   & M0    & 0.28  & --- & ---   & ---   &       &  \\
    5791720 & 3702  & 0.569 & 0.77  & M3    & 4.52  & --- & 0.93  & 1.68$\pm$0.19  & 251   & 80 \\
    5895919 & 3715  & 0.454 & 18.35 & M0    & 0.43  & --- & ---  & ---   &       &  \\
    5953138 & 3499  & 0.36  & 19.12 & M1    & 0     & ---   & --- & ---   &       &  \\
    6205405 & 3723  & 0.48  & ---   & M0    & 0.07  & ---  & 0.01  & ---   &       &  \\
    6224062 & 3707  & 0.452 & 8.6   & M0    & 0.58  & ---  & 0.04  & 1.32$\pm$0.14  & 299   & 94 \\
    6286466 & 3244  & 0.199 & 28.73 & M2    & -0.57 & 0.22 & ---   & ---   &       &  \\
    6436291 & 3687  & 0.563 & 24.08 & M3    & 0.07  & 0.6 & 0.01  & 1.98$\pm$0.17  & 267   & 137 \\
    6526930 & 3772  & 0.5   & 28.93 & M0    & 0.28  & ---  & --- & ---   &       &  \\
    6863726 & 3546  & 0.39  & 21.41 & M0    & 0   & 0.86  & 0.008 & 1.19$\pm$0.09  & 187   & 163 \\
    7417813 & 3661  & 0.556 & 51.43 & M3    & 0.08  & 0.002 & --- & ---  &       &  \\
    7542911 & 3317  & 0.27  & ---   & M2    & 0.18  & 0.004 & ---   & --- &      &  \\
    7949846 & 3530  & 0.38  & 25.17 & M2    & 0.32  & --- & 0.02  & 1.45$\pm$0.14  & 245   & 106 \\
    8143903 & 3465  & 0.34  & 41.63 & M2    & 0.03  & 0.7 & ---   & ---   &       &  \\
    8507979 & 3661  & 0.556 & 1.22  & M3    & 3.78  & 3.48 & 0.62  & 1.25$\pm$0.13  & 206   & 88 \\
    8607728 & 3353  & 0.198 & 24.35 & M3    & 0.58  & --- & 0.015 & 1.45$\pm$0.08  & 433   & 311 \\
    9230350 & 3675  & 0.452 & 16.04 & M0    & -0.14 & 0.81 & 0.001 & 1.12$\pm$0.08  & 234   & 209 \\
    9474589 & 3408  & 0.272 & 22.28 & M1    & 0.31  & --- &  0.01  & 1.12$\pm$0.08  & 263   & 211 \\
    9578520 & 3484  & 0.35  & ---   & M1    & -0.3  & --- & 0.002 & ---   &       &  \\
    9582827 & 3274  & 0.2   & ---   & M3    & 0.1   & ---   & --- & ---   &       &  \\
    9700650 & 3635  & 0.41  & 21.28 & M1    & -0.06 & 0.62 & ---   & ---   &       &  \\
    9705079 & 3277  & 0.239 & 0.54  & M3    & 1.37  & --- & 0.11  & 1.03$\pm$0.17  & 173   & 39 \\
    9730163 & 3327  & 0.189 & ---   & M4    & 3.14  & --- & ---   & ---   &       &  \\
    9946811 & 3684  & 0.44  & ---   & M3    & 2.41  & --- & ---   & ---   &       &  \\
    9955316 & 3682  & 0.413 & 18.09 & M2    & -0.09 & --- & ---  & ---   &       &  \\
    10002261 & 2661  & 0.116 & 12.88 & M3    & 0.77  & --- &  ---   & ---   &       &  \\
    10027247 & 3724  & 0.414 & 17.77 & M0    & -0.11 & --- &  ---   & ---   &       &  \\
    10188460 & 3787  & 0.506 & 13.76 & M2    & 1.44  & --- & 0.05  & 1.30$\pm$0.23  & 246   & 32 \\
    10258179 & 2765  & 0.122 & 32.57 & M3    & -0.13 & --- &0.001 & 1.00$\pm$0.07  & 224   & 214 \\
    10489175 & 3664  & 0.497 & 36.31 & M1    & 0.0     & 0.74  &0.002 & 1.32$\pm$0.11  & 185   & 155 \\
    10598166 & 3488  & 0.398 & ---   & M2    & -0.11 & 0.17 & ---   & ---   &       &  \\
    10713157 & 3749  & 0.45  & 28.57 & M0    & -0.16 & --- &0.001 & 0.91$\pm$0.07 & 185  & 178 \\
    10905746 & 3545  & 0.316 & 18.74 & M2    & 0.58  &0.76 &0.005 & 1.44$\pm$0.07   & 165   & 141 \\
    10961247 & 3720  & 0.436 & 26.55 & M0    & -0.11 & 0.76 &---   & ---   &       &  \\
    11031515 & 3778  & 0.5   & 16.79 & M1    & -0.5  & 0.38& ---   & ---   &       &  \\
    11241109 & 3396  & 0.335 & 55.64 & M2    & 0.57  & 0.94& 0.001 & 1.04$\pm$0.07     & 220   & 209 \\
    11495571 & 3304  & 0.26  & 13.66 & M3    & 2.6   & --- &0.35  & 1.60$\pm$0.31   & 182   & 27 \\
    11521274 & 3482  & 0.35  & 18.48 & M3    & 0.34  & ---  &0.01  & 1.28$\pm$0.08  & 298   & 251 \\
    11615539 & 3766  & 0.509 & 19.7  & M1    & -0.05 & 0.45 &---   & ---   &       &  \\
    11714250 & 3661  & 0.556 & 46.31 & M3    & 0.52  & --- &0.02  & 1.302$\pm$0.10  & 186   & 161 \\
    11763820 & 3471  & 0.38  & 44.65 & M2    & 0.14  & --- &0.001 & ---   &      &  \\
    12505054 & 3521  & 0.35  & 28.44 & M3    & 0.68  & --- &0.02  & 1.52$\pm$0.14  & 210   & 120 \\       

 \end{longtable}
\textbf{Notes.}\\
  $^{a}$Parameter from Huber et al. 2014\\
  $^{b}$Calculated by Lomb-Scargle periodograms\\
  $^{c}$Spectral classification by LAMOST pipeline\\
  $^{d}$H$\alpha$ EW calculated LAMOST pipeline\\
  $^{e}$S-index with SNR $>$ 6\\
  $^{f}$Maximum flare in all observation quarters\\
  $^{g}$Flare frequency distribution\\
  $^{h}$Number of flares in all observation quarters\\
  $^{i}$Number of fitting flares in all observation quarters\\

\clearpage

\end{document}